\documentclass[aps,prb,onecolumn,notitlepage,groupedaddress,showpacs]{revtex4-1}

\usepackage{graphicx}

\begin{document}


\title{Ab initio theory of galvanomagnetic phenomena in ferromagnetic
       metals and disordered alloys}


\author{I. Turek}
\email[]{turek@ipm.cz}
\affiliation{Institute of Physics of Materials,
Academy of Sciences of the Czech Republic,
\v{Z}i\v{z}kova 22, CZ-616 62 Brno, Czech Republic}

\author{J. Kudrnovsk\'y}
\email[]{kudrnov@fzu.cz}
\affiliation{Institute of Physics, 
Academy of Sciences of the Czech Republic,
Na Slovance 2, CZ-182 21 Praha 8, Czech Republic}

\author{V. Drchal}
\email[]{drchal@fzu.cz}
\affiliation{Institute of Physics, 
Academy of Sciences of the Czech Republic,
Na Slovance 2, CZ-182 21 Praha 8, Czech Republic}


\date{\today}

\begin{abstract}
We present an \emph{ab initio} theory of transport quantities of 
metallic ferromagnets developed in the framework of the fully 
relativistic tight-binding linear muffin-tin orbital method.
The approach is based on the Kubo-St\v{r}eda formula for the 
conductivity tensor, on the coherent potential approximation for
random alloys, and on the concept of interatomic electron transport. 
The developed formalism is applied to pure $3d$ transition metals 
(Fe, Co, Ni) and to random Ni-based ferromagnetic alloys (Ni-Fe, 
Ni-Co, Ni-Mn).
High values of the anisotropic magnetoresistance (AMR), found for
Ni-rich alloys, are explained by a negligible disorder in the 
majority spin channel while a change of the sign of the anomalous
Hall effect (AHE) on alloying is interpreted as a band-filling
effect without a direct relation to the high AMR.
The influence of disorder on the AHE in concentrated alloys is
investigated as well.
\end{abstract}


\maketitle


\section{Introduction\label{s_intr}}

Two galvanomagnetic phenomena discovered in the 19th century, 
namely, the anisotropic magnetoresistance (AMR) \cite{r_1857_wt} 
and the anomalous Hall effect (AHE) \cite{r_1881_eh} in
ferromagnets, attract ongoing interest both in basic and applied
physics.
The AMR of bulk systems together with the giant magnetoresistance
of magnetic multilayers plays an important role in magnetic
data storage, \cite{r_2008_smt} whereas the AHE is not fully
understood at present despite the tremendous past and recent 
research activity. \cite{r_2010_nso}
However, the basic origin of both phenomena is well known for 
a long time and it was identified with the simultaneous presence
of spin polarization and spin-orbit (SO) interaction.
Among open problems, one can list, e.g., 
correlations between the anomalous Hall conductivities and
the longitudinal conductivities reported for a number of
systems, \cite{r_2010_nso} or the occurrence of large AMR values 
and the change of sign of the AHE at the same composition (around
15 at.~\% Fe) of random NiFe alloys. \cite{r_1955_js, r_1995_be} 
Let us mention that the AMR and AHE are related, respectively, to
the symmetric and antisymmetric parts of the resistivity tensor
of the solid.
For this reason, an internally consistent theory of all transport
phenomena (resistivity, AMR, AHE, etc.) represents an important
task of condensed-matter physics. 

Existing theoretical techniques for investigation of transport
properties include the Boltzmann equation and the Kubo linear
response theory which represent appropriate tools to describe
electron scattering on impurities or phonons. 
The latter approach uses the Kubo-Greenwood formula \cite{r_1958_dag} 
for longitudinal resistivities and the AMR while the Kubo-St\v{r}eda
formula \cite{r_1982_ps} is a starting point for the AHE. 
\cite{r_2001_cb}
However, the AHE contains not only an extrinsic contribution due to
the electron scattering, but also an intrinsic one that is solely due
to the band structure of an ideal crystal. 
The intrinsic AHE seems to dominate over the extrinsic part in
a number of cases; its values are related to the Berry curvatures
of the Bloch states of the crystal. \cite{r_2010_nso}
These theoretical concepts and techniques have recently been used to
address various problems on a model level, such as, e.g., to describe
quantitatively the AHE and AMR in diluted magnetic semiconductors, 
\cite{r_2002_jnm, r_2006_jsm, r_2009_vks}
or to develop a unified scheme for the AHE of systems with high and
low conductivities. \cite{r_2007_kty, r_2010_ps, r_2010_kst}

Materials-specific \emph{ab initio} theory of galvanomagnetic
phenomena is a straightforward extension of the model-level 
techniques only for the intrinsic AHE of pure magnetic crystals. 
\cite{r_2004_ykm, r_2007_wvy, r_2009_rms} 
Transport properties due to impurity scattering, in particular
residual resistivities, of substitutionally disordered alloys
without the SO interaction have been systematically studied by
means of the Korringa-Kohn-Rostoker (KKR) method and the 
coherent potential approximation (CPA). \cite{r_1985_whb} 
An alternative first-principles technique has recently been
developed within the tight-binding (TB) linear muffin-tin
orbital (LMTO) method. \cite{r_2002_tkd} 
Both approaches agree quantitatively for resistivities of
metallic alloys \cite{r_2002_tkd} and diluted magnetic
semiconductors. \cite{r_2010_sbk}
The AMR of random cubic alloys -- based on diagonal elements
of the conductivity tensor, i.e., on the Kubo-Greenwood
formula -- have been studied in the fully relativistic KKR
method; \cite{r_1996_evb, r_1997_bev, r_2003_veb} 
similar results of the TB-LMTO method with the SO interaction 
included \cite{r_2010_tz} compare well to those of the KKR
method again.
These studies prove that the SO interaction can have a dramatic
effect also on residual resistivities of ferromagnetic alloys
of light ($3d$) elements.

An early attempt \cite{r_1995_be} to calculate the AHE in random
alloys from the Kubo-Greenwood formula using the KKR-CPA method
was shown to be incorrect; \cite{r_2001_cb}
a correct formulation based on the Kubo-St\v{r}eda formula has
appeared only very recently. \cite{r_2010_lke} 
The authors of Ref.~\onlinecite{r_2010_lke} have suggested to 
interpret the coherent part of the anomalous Hall conductivity 
in a random alloy as the intrinsic contribution to the AHE while
the incoherent part -- the co-called vertex corrections -- has been
identified with the extrinsic contribution to the AHE.
Moreover, they have studied the case of dilute alloys and
have shown that the intrinsic contribution exhibits
a well-defined limit for vanishing concentration of impurities
in contrast to a divergence of the extrinsic contribution, 
in reasonable agreement with experimental data for FePd and NiPd
random alloys. 

The purpose of the present paper is to formulate the full 
conductivity tensor of ferromagnetic metals and substitutionally
disordered alloys in the relativistic TB-LMTO-CPA method 
in the atomic sphere approximation (ASA) and to 
illustrate its applicability to the galvanomagnetic phenomena of 
systems containing $3d$ transition-metal elements (Fe, Co, Ni, Mn). 
The paper is organized as follows.
The developed method is presented in Section~\ref{s_meth}.
Section~\ref{ss_rtbl} summarizes the most important relations of
the fully relativistic TB-LMTO formalism relevant for the 
subsequent development of the transport theory, which is given 
in Section~\ref{ss_fct}. 
Technical parts of the derivations are left to Appendices
while details of the numerical procedures employed can be found
in Section~\ref{ss_ind}. 
The calculated results for selected systems and their discussion
are contained in Section~\ref{s_redi}. 
The AMR of Ni-based alloys is discussed in Section~\ref{ss_amr}
while Sections~\ref{ss_ahe1} and \ref{ss_ahe2} are devoted to the
AHE in pure metals and in random alloys, respectively.
The main conclusions of the paper are summarized in
Section~\ref{s_conc}.

\section{Method\label{s_meth}}

\subsection{Relativistic TB-LMTO-ASA method\label{ss_rtbl}}

The Hamiltonian in the orthogonal LMTO representation for a 
ferromagnetic system treated in the fully relativistic LMTO-ASA 
method can be written as \cite{r_1991_slg, r_1996_sdk, r_1997_tdk}
\begin{equation}
H = C + (\sqrt{\Delta})^+ S^0
\left( 1 - \gamma S^0 \right)^{-1} {\sqrt{\Delta}} ,
\label{eq_hortf}
\end{equation}
where the $C$, $\sqrt{\Delta}$ and $\gamma$ are site-diagonal
matrices of potential parameters and the $S^0$ denotes the
matrix of canonical structure constants. 
The form of the Hamiltonian $H$ is similar to the
non-relativistic or non-magnetic cases; 
\cite{r_1975_oka, r_1984_aj, r_1985_ajg, r_1997_tdk} however,
the structure of $H$ is more complicated in the
spin-polarized relativistic case for two reasons. 
\cite{r_1991_slg, r_1996_sdk, r_1997_tdk}
First, the matrices involved in Eq.~(\ref{eq_hortf}) have different
kinds of indices: the site index ${\bf R}$ is combined either
with the usual relativistic index $\Lambda = (\kappa \mu)$, or with
a composed index ${\tilde \Lambda} = (\ell \mu \lambda)$.
Here the quantum number $\mu$ is related to $z$-component of the total
angular momentum, the $\ell$ is related to the orbital angular 
momentum, the non-zero integer $\kappa$ is related in a well-known
manner to the total angular momentum quantum number $j$
($2j+1 = 2|\kappa|$) and to the $\ell$ ($2\ell+1 = |2\kappa + 1|$), 
and the index $\lambda$ labels inequivalent regular solutions of
the single-site problem in each $\ell\mu$-channel. \cite{r_2005_zhs}
The Hamiltonian matrix has thus indices 
$H_{{\bf R}'{\tilde \Lambda}', {\bf R}{\tilde \Lambda}}
= H_{{\bf R}'\ell'\mu'\lambda', {\bf R}\ell\mu\lambda}$ 
while the structure constant matrix has indices
$S^0_{{\bf R}'\Lambda', {\bf R}\Lambda} =
S^0_{{\bf R}'\kappa'\mu', {\bf R}\kappa\mu}$.
Second, the site-diagonal matrices $C$, $\sqrt{\Delta}$ and $\gamma$
are not fully diagonal in the corresponding internal indices, 
($\kappa \mu$) or ($\ell \mu \lambda$), see 
Refs.~\onlinecite{r_1991_slg, r_1996_sdk, r_1997_tdk} for details.
Moreover, the $\sqrt{\Delta}$ is defined as a matrix in the mixed
indices ($\kappa' \mu', \ell \mu \lambda$) which cannot be
understood as a square root of any matrix $\Delta$.
However, for the transport theory developed in Section~\ref{ss_fct},
the detailed structure of the matrices is less important and only 
their general properties are relevant, such as, e.g., the
hermiticity of $C$, $\gamma$ and $S^0$.

The treatment of transport properties of disordered systems requires
the Green's functions (GF).
The basic one -- called the physical GF -- is defined as the resolvent
of the Hamiltonian $H$ (\ref{eq_hortf}):
\begin{equation}
G(z) = \left( z - H \right)^{-1} ,
\label{eq_defgz}
\end{equation}
where $z$ denotes a complex energy variable.
Other GF's are introduced in the TB-LMTO formalism. 
\cite{r_1984_aj, r_1985_ajg, r_1997_tdk}
The matrix of screened structure constants $S^\alpha$ in the 
TB-LMTO representation (superscript $\alpha$) is defined by
\begin{equation}
S^\alpha = S^0 \left( 1 - \alpha S^0 \right)^{-1} ,
\label{eq_scrs}
\end{equation}
where $\alpha$ denotes a site-diagonal matrix of screening constants,
and the site-diagonal matrix of screened potential functions 
$P^\alpha(z)$ is defined as \cite{r_1997_tdk, r_1996_sdk}
\begin{equation}
P^\alpha(z) =  \left[ \sqrt{\Delta} (z - C)^{-1} (\sqrt{\Delta})^+
 + \gamma - \alpha \right]^{-1} .
\label{eq_scrpf}
\end{equation}
The auxiliary GF in the TB-LMTO representation is then defined as
\begin{equation}
g^\alpha(z) = \left[ P^\alpha(z) - S^\alpha \right]^{-1} ,
\label{eq_defga}
\end{equation}
which represents a simpler quantity for theoretical and numerical
treatments than the physical GF $G(z)$, Eq.~(\ref{eq_defgz}).
Both GF's are related to each other by linear rescaling
\begin{equation}
G(z) = \lambda^\alpha(z)
+ \mu^\alpha(z) g^\alpha(z) {\tilde \mu}^\alpha(z) ,
\label{eq_gzvga}
\end{equation}
where the quantities $\lambda^\alpha(z)$, $\mu^\alpha(z)$ and
${\tilde \mu}^\alpha(z)$ denote site-diagonal matrices
\begin{eqnarray}
\lambda^\alpha(z) & = & \mu^\alpha(z) (\gamma - \alpha)
\left[ (\sqrt{\Delta})^+ \right]^{-1} ,
\nonumber\\
\mu^\alpha(z) & = & (\sqrt{\Delta})^{-1}
\left[ 1 + (\alpha - \gamma) P^\alpha(z) \right] ,
\nonumber\\
 {\tilde \mu}^\alpha(z) & = &
\left[ 1 + P^\alpha(z) (\alpha - \gamma) \right]
\left[ (\sqrt{\Delta})^+ \right]^{-1} .
\label{eq_lmtm}
\end{eqnarray}
The relation~(\ref{eq_gzvga}) is indispensable for an efficient
treatment of bulk and layered systems; \cite{r_1997_tdk}
its proof is sketched in Appendix~\ref{app_gf}.

The substitutional disorder in a system on a fixed, non-random
lattice can be best treated in the CPA applied to the auxiliary
GF~(\ref{eq_defga}).
The configurationally averaged $g^\alpha(z)$ is then given by
\cite{r_1997_tdk, r_1990_kd, r_2000_tkd}
\begin{equation} 
\left\langle g^\alpha(z) \right\rangle = {\bar g}^\alpha(z)
= \left[ {\cal P}^\alpha(z) - S^\alpha \right]^{-1} ,
\label{eq_avaux}
\end{equation} 
where the ${\cal P}^\alpha(z)$ denotes the site-diagonal matrix
of the coherent potential functions. 
Their determination from the CPA selfconsistency condition in the
relativistic spin-polarized case as well as details of the
LSDA selfconsistent procedure can be found elsewhere. 
\cite{r_1997_tdk, r_1996_sdk}

\subsection{Full conductivity tensor\label{ss_fct}} 

The conductivity tensor at zero temperature is given according
to the Kubo-St\v{r}eda formula \cite{r_1982_ps, r_2001_cb} as
\begin{eqnarray}
\sigma_{\mu \nu} & = &
\sigma_0\, {\rm Tr} \left\{
  V_\mu \left( G_+ - G_- \right) V_\nu G_- \,
- \, V_\mu G_+ V_\nu \left( G_+ - G_- \right)
 \right.
\nonumber\\
 & & \left. \quad {} +
{\rm i} \left( X_\mu V_\nu - X_\nu V_\mu \right)
\left( G_+ - G_- \right) \right\} ,
\label{eq_contens}
\end{eqnarray} 
where the subscripts $\mu$ and $\nu$ denote indices of Cartesian 
coordinates ($\mu, \nu \in \{ x, y, z \}$), 
the trace (Tr) is taken over all orbitals of the system,
the energy argument of the GF's is omitted since it equals the
Fermi energy $E_{\rm F}$, and we abbreviated
$G_\pm = G(E_{\rm F}\pm {\rm i}0)$.
The symbols $X_\mu$ and $V_\mu$ denote, respectively, the
coordinate and velocity operators. 
The numerical prefactor $\sigma_0$ reflects the units employed 
and the size of the sample; with $\hbar = 1$ assumed here,
it is given by $\sigma_0 = e^2/(4\pi V_0 N)$, where $V_0$ is the
volume of the primitive cell and $N$ is the number of cells in a
big finite crystal with periodic boundary conditions.

In analogy to the non-relativistic formulation of electron
transport, \cite{r_2002_tkd} the coordinate operator $X_\mu$
is represented -- in an orthonormal LMTO basis leading to
the Hamiltonian~(\ref{eq_hortf}) -- by a matrix diagonal
in the (${\bf R}{\tilde \Lambda}$)-index as
\begin{equation}
\left( X_\mu \right)_{{\bf R}'{\tilde \Lambda}', 
{\bf R}{\tilde \Lambda}} = \delta_{{\bf R}'{\bf R}} \, 
\delta_{{\tilde \Lambda}'{\tilde \Lambda}} \, X^\mu_{\bf R} ,
\label{eq_coord}
\end{equation}
where $X^\mu_{\bf R}$ is the $\mu$-th component of the position
vector ${\bf R}$.
The velocity operator $V_\mu$ is then defined as
a quantum-mechanical time derivative of $X_\mu$ ($\hbar = 1$):
\begin{equation}
V_\mu = - {\rm i} \left[ X_\mu , H \right] ,
\label{eq_veloc}
\end{equation}
where $[ A , B ] = AB - BA$ denotes a commutator.
The physical idea behind the simple rule (\ref{eq_coord}) is
an approximation of the true continuous coordinate by its
step-like ``integer'' part that is constant inside each atomic
sphere. 
This leads to a systematic neglect of any intraatomic currents
so that the resulting conductivity $\sigma_{\mu \nu}$ describes
only the net electron motion between neighboring atomic sites.
\cite{r_2002_tkd}
The final simple result for the $X_\mu$ (\ref{eq_coord}) is then
obtained from basic properties of the phi orbitals 
$|\phi_{{\bf R}{\tilde \Lambda}}\rangle$ and phi-dot orbitals
$|{\dot \phi}_{{\bf R}{\tilde \Lambda}}\rangle$ used in the 
definition of the orthonormal LMTO basis. 
\cite{r_1975_oka, r_1984_aj, r_1985_ajg}
In particular, one employs their orthonormality, 
$\langle\phi_{{\bf R}'{\tilde \Lambda}'} | 
\phi_{{\bf R}{\tilde \Lambda}}\rangle
= \delta_{{\bf R}'{\bf R}} 
\delta_{{\tilde \Lambda}'{\tilde \Lambda}}$ and
$\langle\phi_{{\bf R}'{\tilde \Lambda}'} | 
{\dot \phi}_{{\bf R}{\tilde \Lambda}}\rangle = 0$, 
together with a neglect of small quantities 
$\langle{\dot \phi}_{{\bf R}{\tilde \Lambda}'} | 
{\dot \phi}_{{\bf R}{\tilde \Lambda}}\rangle$ which is consistent
with the second-order accuracy of the LMTO 
Hamiltonian~(\ref{eq_hortf}).

For practical calculations, one can recast the original form
of the conductivity tensor $\sigma_{\mu \nu}$ (\ref{eq_contens})
into a more suitable version in which the velocities
$V_\mu$ (\ref{eq_veloc}) are replaced by effective velocities
$v^\alpha_\mu$ defined by
\begin{equation}
v^\alpha_\mu = - {\rm i} \left[ X_\mu , S^\alpha \right] ,
\label{eq_veloce}
\end{equation}
and in which the physical GF's $G_\pm$ are replaced by the
auxiliary GF's $g^\alpha_\pm = g^\alpha(E_{\rm F}\pm {\rm i}0)$,
see Eq.~(\ref{eq_defga}).
This transformation rests on two relations.
The first one connects both velocity operators by
\begin{equation}
V_\mu = (\sqrt{\Delta})^+ (F^\alpha)^{-1} \,
v^\alpha_\mu \, [( F^\alpha )^+]^{-1} \sqrt{\Delta} ,
\label{eq_velop}
\end{equation}
where
\begin{equation}
F^\alpha = 1 + S^\alpha (\alpha - \gamma)
\label{eq_falpha}
\end{equation}
and, consequently, $(F^\alpha)^+ = 1+(\alpha - \gamma)S^\alpha$.
The velocity relation (\ref{eq_velop}) can easily be obtained
from the explicit form of
$H$ (\ref{eq_hortf}) and from an identity
$S^0 ( 1 - \gamma S^0 )^{-1} = (F^\alpha)^{-1} S^\alpha
= S^\alpha [(F^\alpha)^+]^{-1}$, which is a direct consequence
of the screening transformation (\ref{eq_scrs}).
Note that the coordinate operator $X_\mu$ in the definition
of the effective velocity $v^\alpha_\mu$ (\ref{eq_veloce}) has
to be understood as a diagonal matrix in the 
(${\bf R}\Lambda$)-index, i.e.,
\begin{equation}
\left( X_\mu \right)_{{\bf R}'\Lambda', {\bf R}\Lambda} =
\delta_{{\bf R}'{\bf R}} \, \delta_{\Lambda'\Lambda} \, 
X^\mu_{\bf R} ,
\label{eq_coorde}
\end{equation}
and that both operators $X_\mu$, Eqs.~(\ref{eq_coord})
and (\ref{eq_coorde}), commute with all other site-diagonal
operators ($C$, $\sqrt{\Delta}$, $\gamma$, $\alpha$).
The second relation connects the physical and auxiliary GF's by
\begin{equation}
G(z) = (\sqrt{\Delta})^{-1} (F^\alpha)^+
\left[ (\alpha - \gamma) + g^\alpha(z) F^\alpha \right]
[(\sqrt{\Delta})^+]^{-1} ,
\label{eq_agvga}
\end{equation}
which represents a complementary relation to Eq.~(\ref{eq_gzvga})
and which is proved in a similar way in Appendix \ref{app_gf}.
The use of Eqs.~(\ref{eq_velop}) and (\ref{eq_agvga}) in 
the conductivity tensor (\ref{eq_contens}) yields another form of
the latter, namely,
\begin{eqnarray} 
\sigma_{\mu \nu} & = &
 \sigma_0\, {\rm Tr} \left\{
  v^\alpha_\mu \left( g^\alpha_+ - g^\alpha_- \right)
    v^\alpha_\nu g^\alpha_- \,
- \, v^\alpha_\mu g^\alpha_+ v^\alpha_\nu
 \left( g^\alpha_+ - g^\alpha_- \right) \right.
\nonumber\\
 & & \left. \quad {} +
{\rm i} \left( X_\mu v^\alpha_\nu - X_\nu v^\alpha_\mu \right)
\left( g^\alpha_+ - g^\alpha_- \right) \right\} .
\label{eq_contenf}
\end{eqnarray}
The proof of equivalence of both expressions for the 
$\sigma_{\mu \nu}$ is given in Appendix \ref{app_ct}.

The similarity of both conductivity formulas, Eqs.~(\ref{eq_contens})
and (\ref{eq_contenf}), is obvious. 
However, the latter one is better suited for configuration averaging
since the effective velocities $v_\mu^\alpha$ (\ref{eq_veloce}) are
non-random operators in contrast to the original velocities $V_\mu$
(\ref{eq_veloc}) and to the velocities in the KKR-CPA method.
The CPA-average of the $\sigma_{\mu \nu}$ (\ref{eq_contenf}) can 
thus be reduced to averages of the auxiliary GF's (\ref{eq_avaux})
and to the standard form of vertex corrections for the CPA-average
of products $g^\alpha(z_1) \otimes g^\alpha(z_2)$. \cite{r_1969_bv}
The same transformation was achieved in our previous study
\cite{r_2002_tkd} for longitudinal conductivities 
within the Kubo-Greenwood formula \cite{r_1958_dag} and
the non-relativistic TB-LMTO method.
The derived result, Eq.~(\ref{eq_contenf}), represents an extension
in two directions: to spin-polarized relativistic systems and to
the full conductivity tensor, including the anomalous Hall
conductivities.

\subsection{Implementation and numerical details\label{ss_ind}} 

The fully relativistic theory developed in Section~\ref{ss_fct} has
been applied to random binary alloys of $3d$ transition metals 
(Mn, Fe, Co, Ni), where the effects of the SO interaction are rather
weak as compared to the alloy bandwidths and exchange splittings.
For this reason, most of the results shown in the next section were
calculated by means of a simplified version of the theory,
in which the SO coupling was included as an on-site perturbation
term of the $\xi {\bf L} \cdot {\bf S}$ form to the LMTO Hamiltonian
in the scalar-relativistic approximation (SRA).
\cite{r_1975_oka, r_2008_tdk}
This SRA+SO approach proved to yield results in surprisingly good
quantitative agreement with results of fully relativistic LMTO and
KKR techniques, both for magnetic moments and densities of states
\cite{r_2008_tdk} and for longitudinal resistivities \cite{r_2010_tz}
of $3d$ transition-metal systems. 
In a few selected cases, results of the fully relativistic Dirac
(FRD) theory \cite{r_1996_sdk} will be shown for a direct comparison
of both approaches. 

The employed LMTO basis comprised $s$, $p$, and $d$-type orbitals,
and the LSDA selfconsistency was achieved with the Vosko-Wilk-Nusair
exchange-correlation potential \cite{r_1980_vwn} and for magnetization
pointing along $z$-axis of the alloy cubic structures.
More details about the ground-state electronic structure calculations
were published elsewhere. \cite{r_1997_tdk, r_2008_tdk, r_1996_sdk}
The numerical prefactor $\sigma_0$ used in the final conductivity 
formula, Eq.~(\ref{eq_contenf}), is given by $\sigma_0 = 
e^2/(4\pi \hbar V_0 N)$, where $N$ is the number of ${\bf k}$ vectors
sampling the whole Brillouin zone (BZ). \cite{r_2002_tkd}
The terms bilinear in the auxiliary GF's $g_\pm$ in 
Eq.~(\ref{eq_contenf}) were configurationally averaged in the CPA
with the corresponding vertex corrections included, \cite{r_2006_ctk}
whereas the term linear in the $g_\pm$ has been omitted here for 
symmetry reasons \cite{r_2010_lke} (it vanishes identically due to
the inversion symmetry of the Bravais lattice).
In a general case, the latter term in Eq.~(\ref{eq_contenf}) is
essentially equivalent to the so-called Fermi-sea contribution to 
the AHE, which can be neglected in metallic 
systems. \cite{r_2010_nhk}

\begin{figure}
\includegraphics[width=0.45\textwidth]{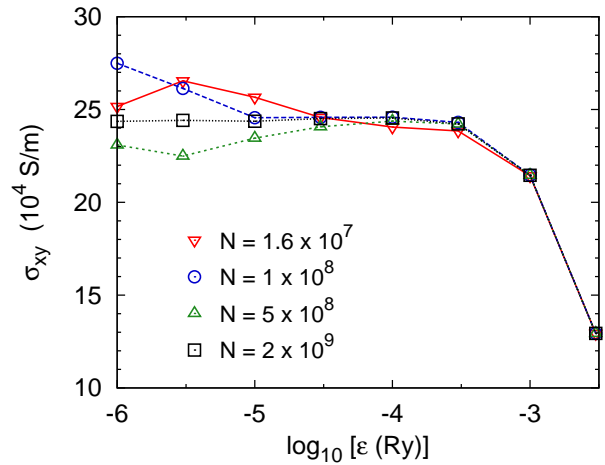}
\caption{(Color online) 
The anomalous Hall conductivity $\sigma_{xy}$ of fcc Ni versus 
the imaginary part of energy $\varepsilon$ for different
numbers $N$ of ${\bf k}$ vectors in the full BZ. 
The converged value obtained by methods based
on the Berry curvature is $\sigma_{xy}=22.0\times 10^4$~S/m,
see Ref.~\onlinecite{r_2007_wvy} and Table~\ref{t_ahc} below.
\label{f_ni_conv}}
\end{figure}

Particular attention was paid to convergence behavior of the
conductivities with respect to the number $N$ of sampling ${\bf k}$
points.
We found that large $N$'s were necessary and that energy arguments
of the GF's had to contain a non-zero imaginary part, 
$z = E_{\rm F} \pm {\rm i}\varepsilon$.
The quantity $\varepsilon$ should be as small as possible for
purely theoretical values of $\sigma_{\mu \nu}$; however,
for a comparison with experiment, finite values of $\varepsilon$ can
also be used in order to simulate additional scattering on phonons,
magnons and defects in real samples. \cite{r_2009_ktt}
Figure~\ref{f_ni_conv} displays the anomalous Hall conductivity 
$\sigma_{xy}$ of fcc Ni as a function of $\varepsilon$ and $N$. 
One can see that converged values of $\sigma_{xy}$ for small values
of $\varepsilon$ require at least $N \sim 10^8$ sampling points in
the full BZ.
In this study, we used $\varepsilon = 10^{-5}$~Ry and 
$N \approx 4\times 10^8$ for bcc Fe while
$N \approx 9\times 10^8$ were used for fcc Co and Ni-based systems.

\section{Results and discussion\label{s_redi}}

For cubic systems with magnetization along $z$-axis, the 
conductivity tensor has three independent non-zero matrix elements,
namely, $\sigma_{xx}=\sigma_{yy}$, $\sigma_{zz}$, and
$\sigma_{xy}=-\sigma_{yx}$; the same structure is found for the
resistivity tensor $\rho_{\mu \nu}$.
The AHE is obtained quantitatively from the non-diagonal elements,
$\sigma_{xy}$ or $\rho_{xy}$, while the diagonal elements define
the isotropic resistivity $\rho = (2 \rho_{xx} + \rho_{zz})/3$ 
and the AMR ratio $\Delta \rho / \rho$, where 
$\Delta \rho = \rho_{zz} - \rho_{xx}$.

\subsection{AMR in random fcc Ni-based alloys\label{ss_amr}}

A brief summary of the present TB-LMTO results and their
comparison to available experimental data is shown in 
Table \ref{t_nib} for three Ni-based ferromagnetic alloys with
the same average number of electrons per atom ${\bar Z} = 27.7$.
One can see that the resistivities of the Ni-Co and Ni-Fe
alloys are smaller by a factor of two to four as compared
to experimental values. \cite{r_1977_jcf, r_2010_tpr}
Similar underestimation of the resistivity values has been obtained
by the relativistic KKR-CPA method \cite{r_1995_be, r_1997_bev} 
while the scattering theory combined with a supercell TB-LMTO
approach \cite{r_2010_skb} yields resistivities that are only
$\sim 20\%$ smaller than the measured ones.
The calculated AMR values of Ni-Co and Ni-Fe values exceed slightly
the measured ones while the opposite discrepancy is found for the
ferromagnetic Ni-Mn system: for a Ni$_{0.97}$Mn$_{0.03}$ alloy, the
measured AMR amounts to $7.3\%$, \cite{r_1975_dm} whereas the present
calculation yields the AMR of $3.3\%$ and $2.4\%$ in the FRD 
and the SRA+SO approach, respectively. 

\begin{table}
\caption{Comparison of the calculated and measured transport
quantities for three fcc Ni-based alloys with the average number of
electrons per atom ${\bar Z} = 27.7$: the isotropic resistivity 
$\rho$ (in $\mu \Omega$cm), the AMR ratio and the anomalous Hall 
conductivity $\sigma_{xy}$ (in kS/m).
The displayed calculated values correspond to the FRD approach;
in parentheses, results of the SRA+SO approach are shown.
The experimental values are taken from Ref.~\onlinecite{r_1977_jcf}. 
\label{t_nib}}
\begin{ruledtabular}
\begin{tabular}{lccc}
alloy         & $\rho$ & AMR & $\sigma_{xy}$ \\
\hline
Ni$_{0.7}$Co$_{0.3}$ calc. & 1.06(0.70) & 0.47(0.42) & $-88$($-74$) \\
\qquad \qquad \quad exp. & 2.8 & 0.3 & - \\
\hline
Ni$_{0.85}$Fe$_{0.15}$ calc. & 2.21(1.57) & 0.29(0.24) & $-40$($-29$) \\
\qquad \qquad \quad exp. & 4.2 & 0.18 & - \\
\hline
Ni$_{0.9}$Mn$_{0.1}$ calc. & 22.8(23.6) & 0.009(0.004) & $65$($54$) \\
\end{tabular}
\end{ruledtabular}
\end{table}

\begin{figure}
\includegraphics[width=0.45\textwidth]{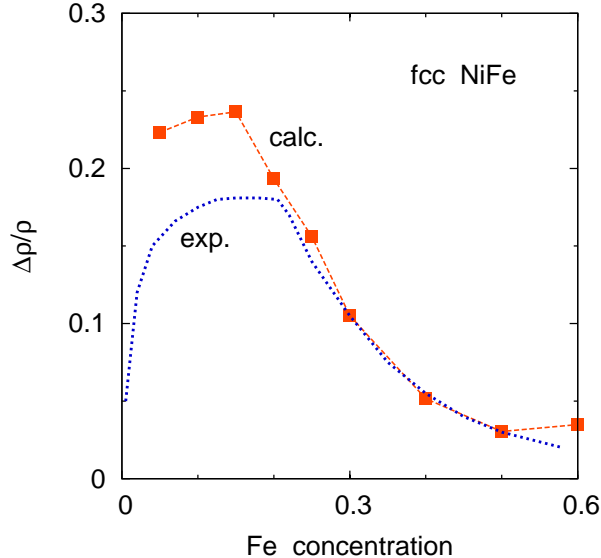}
\caption{(Color online) 
The calculated values of the anisotropic magnetoresistance
of random fcc Ni$_{1-x}$Fe$_x$ alloy as a function of Fe 
concentration and their comparison to experimental 
values. \cite{r_1977_jcf}
\label{f_nife_amr}}
\end{figure}

The first application of the developed scheme has been done for
the AMR of fcc Ni-based alloys.
A preliminary study of the Ni-Co and Ni-Cu alloy systems including 
a comparison with the results of the KKR method was published
elsewhere; \cite{r_2010_tz} here we show the AMR for the fcc
Ni$_{1-x}$Fe$_x$ alloy calculated in the SRA+SO approach, 
see Fig.~\ref{f_nife_amr}. 
One can see relatively high values of the AMR for the Ni-rich
alloys with the maximum AMR ratio slightly exceeding 20\%. 
Similar trends and values were obtained in the fully relativistic
KKR method \cite{r_1995_be, r_2003_kps} and in 
experiments. \cite{r_1977_jcf}

The high AMR values for the Ni-rich Ni-Fe alloys and even
higher theoretical \cite{r_1997_bev, r_2010_tz} as well as 
experimental \cite{r_1977_jcf, r_2010_tpr} values for the Ni-rich
Ni-Co alloys deserve attention and call for explanation. 
Existing theoretical approaches to the AMR in metals and alloys
are based on inclusion of the SO interaction in the lowest-order
perturbation expansion. \cite{r_1951_js, r_1977_jcf, r_2012_kth}
However, the validity of such schemes for the Ni-rich alloys is
limited due to the well-known fact that the SO interaction increases
the resistivity by a factor of two or more as compared to the
resistivity within the two current model. \cite{r_1997_bev, 
r_2010_tz}
In order to reveal the possible origin of the high AMR in these
systems, we considered also the fcc Ni-Mn alloy with
a ferromagnetic order in the whole concentration range studied
(which represents the true ground state only for Mn concentrations
below 15 at.\% Mn \cite{r_1993_ad, r_2008_kdb})
and an artificial fcc Ni-Fe(*) alloy in which the majority spin
(spin-up) potential of Fe atoms was replaced by that of Ni atoms
leading thus to a system with no spin-up disorder.
The real Ni-Co and Ni-Fe alloys are featured by a very weak
disorder in the majority spin channel; \cite{r_1997_bev}
the four ferromagnetic alloys considered in the present study
form a set in which the spin-up disorder decreases along
the sequence Ni-Mn, Ni-Fe, Ni-Co, and Ni-Fe(*).

\begin{figure}
\includegraphics[width=0.45\textwidth]{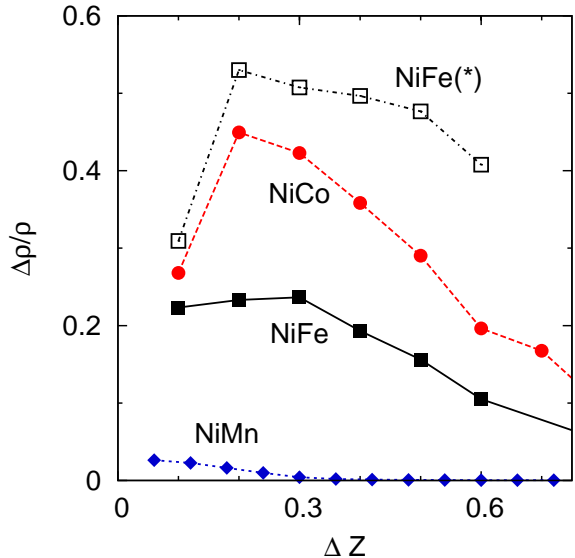}
\caption{(Color online) 
The calculated values of the anisotropic magnetoresistance
of random fcc Ni-based alloys as functions of the valence
charge difference $\Delta Z$.
\label{f_nibase_amr}}
\end{figure}

The resulting AMR values of the four systems are plotted in 
Fig.~\ref{f_nibase_amr} as functions of the change of the
number of valence electrons $\Delta Z$ due to alloying.
The latter quantity is defined as 
$\Delta Z(x) = x$ for the Ni$_{1-x}$Co$_x$ alloy, 
$\Delta Z(x) = 2x$ for the Ni$_{1-x}$Fe$_x$ alloy, and
$\Delta Z(x) = 3x$ for the Ni$_{1-x}$Mn$_x$ alloy.
The calculated AMR increases along the sequence
Ni-Mn, Ni-Fe, Ni-Co, Ni-Fe(*), in qualitative agreement
with experiment for the real alloys. \cite{r_1977_jcf}
The obtained trend indicates that the AMR ratios are directly
related to the disorder strength in the majority spin channel,
with the highest AMR ratios obtained for alloys with negligible
spin-up scattering. 
This conclusion does not depend on the particular approach
(SRA+SO, FRD) employed in the calculation, as can be seen from a
comparison of results for $\Delta Z = 0.3$, see Table \ref{t_nib}.

\subsection{AHE in elemental ferromagnets\label{ss_ahe1}}

The calculated anomalous Hall conductivities $\sigma_{xy}$ of
three cubic $3d$ transition metals are listed in Table~\ref{t_ahc}
together with other theoretical and experimental results.
The Co was treated here in the fcc structure, which represents 
the natural cubic approximation to the ground-state hexagonal 
close-packed structure. \cite{r_2009_rms} 
It should be noted that our sign convention for the $\sigma_{xy}$
differs from that adopted in the previous 
studies; \cite{r_2004_ykm, r_2007_wvy, r_2009_rms, r_2010_lke}
all signs in the table are thus taken consistently with our 
convention.

\begin{table}
\caption{The calculated and experimental values of the
anomalous Hall conductivity $\sigma_{xy}$ (in kS/m) for
$3d$ transition-metal ferromagnets.
The displayed values obtained in this work correspond to the FRD 
approach; in parentheses, results of the SRA+SO approach are shown.
\label{t_ahc}}
\begin{ruledtabular}
\begin{tabular}{lccc}
         & bcc Fe & fcc Co & fcc Ni \\
\hline
this work & $-65$($-55$) & $-36$($-34$) & $241$($243$) \\
Berry curvature & $-75^a$ & $-25^b$ & $220^c$ \\
KKR method & $-64^d$ & & $164^d$ \\
experiment & $-103^e$ & & $65^f$ \\
\end{tabular}
\end{ruledtabular}
$^a$ Reference \onlinecite{r_2004_ykm}. 
$^b$ Reference \onlinecite{r_2009_rms}. 
$^c$ Reference \onlinecite{r_2007_wvy}. 
$^d$ Reference \onlinecite{r_2010_lke}. 
$^e$ Reference \onlinecite{r_1967_pnd}. 
$^f$ Reference \onlinecite{r_1961_jml}. 
\end{table}

One can see that the results of the TB-LMTO method agree reasonably
well with the data based on the Berry 
curvature \cite{r_2004_ykm, r_2007_wvy, r_2009_rms} and on the 
fully relativistic KKR Green's-function technique. \cite{r_2010_lke}
The biggest discrepancy in Table~\ref{t_ahc}, related to the
measured and calculated AHE of fcc Ni, might reflect effects of
strong correlations on the electronic structure. \cite{r_2011_fg}

\subsection{AHE in random fcc Ni-based alloys\label{ss_ahe2}}

\begin{figure}
\includegraphics[width=0.45\textwidth]{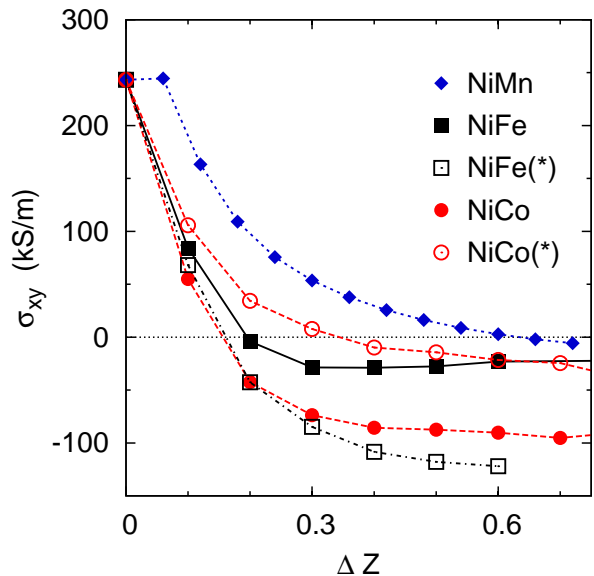}
\caption{(Color online) 
The calculated values of the anomalous Hall conductivity 
$\sigma_{xy}$ of random fcc Ni-based alloys as functions of the
valence charge difference $\Delta Z$.
\label{f_nibase_ahe}}
\end{figure}

The effect of alloying on the AHE has been studied for the random
fcc Ni-based alloys discussed in Section~\ref{ss_amr}.
In order to assess the role of chemical disorder on the AHE,
the Ni$_{1-x}$Co$_x$ alloy has also been treated in
a simple virtual crystal approximation (VCA) in which the alloy
constituents are replaced by an effective atom with the atomic 
number $Z(x) = 28 - x$; the resulting system is denoted as
Ni-Co(*).

The concentration trends of the $\sigma_{xy}$, calculated in the
SRA+SO approach, are summarized in Fig.~\ref{f_nibase_ahe}.
We note that the dilute limit of the alloys, where the $\sigma_{xy}$
diverges due to the diverging incoherent contribution,
\cite{r_2010_lke} has not been studied here.
All studied systems exhibit a change of the sign of the AHE due to
alloying.
This can be understood as a consequence of band filling and of the
opposite signs of the $\sigma_{xy}$ for pure fcc Co and Ni, see
Table~\ref{t_ahc}.
A similar sign change has been observed a long time ago for Ni-rich
Ni-Fe and Ni-Co alloys. \cite{r_1955_js, r_1964_lb, r_1964_kck}
This feature was related to the occurrence of the high AMR values
encountered in these systems with roughly the same composition.
\cite{r_1964_lb, r_1995_be}
However, our calculated data for the ferromagnetic Ni-Mn alloy
witness, that the sign change of the AHE can be obtained also for
a system with much smaller AMR values.

Let us discuss finally the role of disorder on the AHE from
a comparison of the calculated $\sigma_{xy}$ for the Ni-Fe and
Ni-Fe(*) systems as well as for the Ni-Co and Ni-Co(*) alloys.
As one can see from Fig.~\ref{f_nibase_ahe}, the neglect of the
disorder in the majority spin channel (Ni-Fe) or in both spin
channels (Ni-Co) has a strong influence on the resulting
$\sigma_{xy}$.
This analysis indicates that results of simplified treatments
of alloying, such as, e.g., in Ref.~\onlinecite{r_2011_zbm}, 
should be taken with caution.
Moreover, the $\sigma^{\rm VCA}_{xy}$ for the Ni-Co(*) alloy
differs significantly both from the total $\sigma_{xy}$ for the
corresponding Ni-Co alloy in the CPA and from its coherent part,
$\sigma^{\rm coh}_{xy}$.
For the equiconcentration Ni-Co alloy, we get:
$\sigma^{\rm VCA}_{xy} = -14$ kS/m, $\sigma_{xy} = - 88$ kS/m,
and $\sigma^{\rm coh}_{xy} = -93$ kS/m. 
This example proves that the coherent part of the anomalous Hall
conductivity of a real concentrated alloy is not related directly
to that of the corresponding effective crystal, in contrast to the
case of the dilute alloys. \cite{r_2010_lke}

\section{Conclusions\label{s_conc}}

We have reformulated the Kubo-St\v{r}eda expression
for the conductivity tensor \cite{r_1982_ps, r_2001_cb} 
in the fully relativistic TB-LMTO-CPA theory. \cite{r_1996_sdk}
This extends the applicability of our previous transport 
formalism \cite{r_2002_tkd} to all elements of the conductivity
tensor, indispensable for the anomalous Hall effect.
The effect of the spin-orbit interaction has been included and
implemented numerically in the Dirac four-component formalism
as well as in a simple perturbative manner.
The former approach is inevitable for systems with heavy elements
while the latter scheme is advantageous for interpretation of the
results in terms of the underlying electronic structure owing to
the non-relativistic (or scalar-relativistic) basis set of orbitals.
First applications to galvanomagnetic phenomena in pure metals
(Fe, Co, Ni) and in ferromagnetic Ni-based alloys yield results in
reasonable agreement with other techniques. 
The performed calculations for the real and artificial alloy systems
clarified some aspects of the anisotropic magnetoresistance
and the anomalous Hall effect and their possible interrelation. 
The developed technique has already been generalized and applied to
alloy structures with several sublattices and different degree of
atomic ordering. \cite{r_2011_kdk}
Further questions, such as, e.g., the influence of complex magnetic
orders on the transport properties in systems with competing ferro-
and antiferromagnetic interactions, remain a task for future.

\begin{acknowledgments}
The authors acknowledge financial support by the Czech Science
Foundation (Grant No.\ P204/11/1228).
\end{acknowledgments}


\appendix
\section{Relations between the Green's functions\label{app_gf}}

The proof of Eqs.~(\ref{eq_gzvga}) and (\ref{eq_agvga}) is based on
a general identity valid for matrices ${\hat x}$, $x$, ${\hat y}$, 
$y$, and $f$, that are coupled by
\begin{eqnarray}
{\hat x} = x (1+fx)^{-1} = (1+xf)^{-1} x , \qquad
& &  
x = {\hat x} (1-f{\hat x})^{-1} = (1-{\hat x}f)^{-1} {\hat x} , 
\nonumber\\
{\hat y} = y (1+fy)^{-1} = (1+yf)^{-1} y , \qquad
& &  
y = {\hat y} (1-f{\hat y})^{-1} = (1-{\hat y}f)^{-1} {\hat y} ,
\label{eq_defxyf}
\end{eqnarray}
if corresponding inverses exist.
This yields relations
\begin{eqnarray}
{\hat x}-{\hat y} & = & (1+xf)^{-1} (x-y) (1+fy)^{-1} , 
\nonumber\\
\left( {\hat x}-{\hat y} \right)^{-1} & = & 
(1+fy) (x-y)^{-1} (1+xf) ,
\label{eq_xmy}
\end{eqnarray}
and the general identity
\begin{equation}
\left( {\hat x}-{\hat y} \right)^{-1} =  
- (1+fx)f + (1+fx) (x-y)^{-1} (1+xf) . 
\label{eq_auxxy}
\end{equation}

This identity can be used for 
${\hat x} = [(\sqrt{\Delta})^+]^{-1} (z-C) (\sqrt{\Delta})^{-1}$, 
${\hat y} = S^0 ( 1 - \gamma S^0 )^{-1}$, and
$f = \alpha - \gamma$, so that $x = P^\alpha(z)$ and $y = S^\alpha$.
Substitution into Eq.~(\ref{eq_auxxy}) and subsequent multiplication
by $(\sqrt{\Delta})^{-1}$ from the left and by
$[(\sqrt{\Delta})^+]^{-1}$ from the right leads to the
relation~(\ref{eq_gzvga}) with the matrices $\lambda^\alpha(z)$,
$\mu^\alpha(z)$ and ${\tilde \mu}^\alpha(z)$ given by 
Eq.~(\ref{eq_lmtm}). 
Similarly, the identity~(\ref{eq_auxxy}) can be applied to 
${\hat y} = [(\sqrt{\Delta})^+]^{-1} (z-C) (\sqrt{\Delta})^{-1}$,
${\hat x} = S^0 ( 1 - \gamma S^0 )^{-1}$, and
$f = \alpha - \gamma$, so that $y = P^\alpha(z)$ and $x = S^\alpha$.
This yields the complementary relation between the Green's functions,
Eq.~(\ref{eq_agvga}).

\section{Equivalence of expressions for the conductivity
           tensor\label{app_ct}}

For the proof of equivalence of Eqs.~(\ref{eq_contens}) and
(\ref{eq_contenf}), we drop the superscript $\alpha$ at
$v^\alpha_\mu$, $g^\alpha_\pm$, $S^\alpha$ and $F^\alpha$,
and abbreviate $a = \alpha - \gamma$.
The definition (\ref{eq_falpha}) is thus written as
\begin{equation}
F = 1 + S a ,
\label{eq_abbrf}
\end{equation}
and the two relations (\ref{eq_velop}) and (\ref{eq_agvga}) 
are now written as
\begin{equation}
V_\mu = (\sqrt{\Delta})^+ F^{-1} v_\mu (F^+)^{-1} \sqrt{\Delta} 
, \qquad
G_\pm = (\sqrt{\Delta})^{-1} F^+ ( a + g_\pm F )
 [(\sqrt{\Delta})^+]^{-1} .
\label{eq_vgabb}
\end{equation}
Substitution of Eq.~(\ref{eq_vgabb}) into the original
Eq.~(\ref{eq_contens}) yields after trivial modifications 
\begin{eqnarray}
\sigma_{\mu \nu} & = & 
\sigma_0\, {\rm Tr} \left\{ 
F^{-1} v_\mu \left( g_+ - g_- \right) 
v_\nu \left( a + g_- F \right) -
v_\mu \left( a + g_+ F \right)
F^{-1} v_\nu \left( g_+ - g_- \right) 
\right. 
\nonumber\\
  & & \left. \ \, {} + {\rm i} \left( 
X_\mu F^{-1} v_\nu - X_\nu F^{-1} v_\mu 
\right) \left( g_+ - g_- \right) F 
\right\} \ = \ 
\sigma^{(1)}_{\mu \nu} + \sigma^{(2)}_{\mu \nu} , 
\label{eq_conten_b}
\end{eqnarray}
where the $\sigma^{(1)}_{\mu \nu}$ and $\sigma^{(2)}_{\mu \nu}$ 
comprise, respectively, all terms bilinear and linear in
$g_\pm$.
The first contribution is obvious,
\begin{equation}
\sigma^{(1)}_{\mu \nu} =  
\sigma_0\, {\rm Tr} \left\{ 
   v_\mu \left( g_+ - g_- \right) v_\nu g_- \,
 - \, v_\mu g_+ v_\nu \left( g_+ - g_- \right) 
\right\} ,
\label{eq_conten_1}
\end{equation} 
while the second contribution is more complicated. 
It has a form 
\begin{equation}
\sigma^{(2)}_{\mu \nu} = \sigma_0\, {\rm Tr} 
\left\{ N_{\mu \nu} \left( g_+ - g_- \right) \right\} ,
\label{eq_conten_2w}
\end{equation} 
where the $N_{\mu \nu}$ can be written in terms of the effective
velocities $v_\mu$ (\ref{eq_veloce}) as 
\begin{eqnarray}
N_{\mu \nu} & = & 
v_\nu a F^{-1} v_\mu - v_\mu a F^{-1} v_\nu 
 + {\rm i} F \left( 
X_\mu F^{-1} v_\nu - X_\nu F^{-1} v_\mu 
\right)
\nonumber\\
 & = & {\rm i} \left( [ X_\mu , S ] a 
+ F X_\mu \right) F^{-1} v_\nu 
     - {\rm i} \left( [ X_\nu , S ] a 
+ F X_\nu \right) F^{-1} v_\mu . 
\qquad
\label{eq_conten_ns}
\end{eqnarray}
Here, the last two brackets can be modified using 
Eq.~(\ref{eq_abbrf}) and $[ X_\mu , a ] = 0$, so that 
$[ X_\mu , F ] = [ X_\mu , S ] a$, hence
$[ X_\mu , S ] a + F X_\mu = X_\mu F$ and   
Eq.~(\ref{eq_conten_ns}) reduces to
\begin{equation}
N_{\mu \nu} = {\rm i} \left( 
X_\mu v_\nu - X_\nu v_\mu \right) .
\label{eq_conten_nf}
\end{equation}
Consequently, the second contribution in Eq.~(\ref{eq_conten_b}) 
simplifies to
\begin{equation}
\sigma^{(2)}_{\mu \nu} = \sigma_0\, {\rm Tr} 
\left\{ {\rm i} \left( 
X_\mu v_\nu - X_\nu v_\mu \right)  
\left( g_+ - g_- \right) \right\} .
\label{eq_conten_2f}
\end{equation} 
The sum of Eq.~(\ref{eq_conten_1}) and Eq.~(\ref{eq_conten_2f})
is now identical to the transformed expression for the
$\sigma_{\mu \nu}$ (\ref{eq_contenf}). 


\providecommand{\noopsort}[1]{}\providecommand{\singleletter}[1]{#1}%

\end{document}